\documentclass{ws-procs975x65}

\begin{document}

\title{On the mass of the gravitational lenses in LMC}

\author{L. MANCINI\footnote{\uppercase{L.M.} is partially
supported by the \uppercase{S}wiss \uppercase{N}ational
\uppercase{S}cience \uppercase{F}oundation.} \hspace{0.1mm} and G.
SCARPETTA\footnote{\uppercase{G.S.} is partially supported by the
\uppercase{I}stituto \uppercase{N}azionale di \uppercase{F}isica
\uppercase{N}ucleare, sez. \uppercase{N}apoli,
\uppercase{I}taly.}}

\address{Dipartimento di Fisica,
           Universit\`{a} di Salerno, I-84081 Baronissi (SA),
           Italy\\}

\author{S. CALCHI NOVATI\footnote{\uppercase{S.C.N.} is
supported by the \uppercase{S}wiss \uppercase{N}ational
\uppercase{S}cience \uppercase{F}oundation and the
\uppercase{T}omalla \uppercase{F}oundation.} \hspace{0.1mm} and
PH. JETZER}

\address{Institut f\"{u}r Theoretische Physik der
           Universit\"{a}t Z\"{u}rich, CH-8057 Z\"{u}rich,
           Switzerland}

%

\maketitle

\abstracts{In the self--lensing framework, we estimate the modal
values of the mass of the gravitational lenses found by the MACHO
collaboration towards the Large Magellanic Cloud (LMC). Our
results suggest that only the events located near the center can
be identified as a low mass star population belonging to the LMC
disk or bar components.}
\section{Microlensing towards the Large Magellanic Cloud}
The microlensing surveys towards the Large Magellanic Cloud (LMC)
have demonstrated the existence of compact objects that act as
gravitational lenses somewhere between us and LMC. Besides the 4
events found by the EROS2 group [1], the MACHO Project has
detected 16 events and its maximum likelihood estimate of the mass
of the lensing objects is $\approx 0.5$ M$_{\odot}$, while the
fraction $f$ of dark matter in form of MACHOs in the Galactic halo
is estimated to be $\sim 20\%$ [2]. However, the interpretation of
these data is a matter of controversy and other hypothesis have
been proposed: the possibility that the lenses are located in LMC
itself instead as in the Galactic halo was claimed by Sahu [3]; a
warp of the Milky Way (MW) disk, which covers the line of sight
towards LMC, has been proposed to support a lens disk population
[4]; Zhao [5] has suggested instead that the debris torn from the
LMC by tidal forces may be a source of MACHOs; a non planar
geometry of LMC, i.e. a misalignment of the bar from the the disk,
has been proposed by Zhao \& Evans [6]; other authors have
considered LMC components fatter than is conventional, with
material extending to scale heights of $\sim 6$ kpc above the
plane of the LMC disk [7], as is suggested by Weinberg [8]
numerical simulations of the evolution of the LMC in the tidal
field of the MW. The analysis of Jetzer et al. [9] has shown that
possibly the observed events are distributed among different
galactic components (disk, spheroid, galactic halo, LMC halo and
self-lensing). This means that the lenses do not belong all to the
same population and their astrophysical features can differ deeply
one another. The microlensing surveys towards the Small Magellanic
Cloud have detected very few events, which did not help to clarify
the problem as it was hoped.
%
\section{A new picture of the Large Magellanic Cloud}
A new accurate description of the LMC geometry and dynamics was
recently proposed by van der Marel et al. [10], thanks to the
DENIS and 2MASS surveys. From their study emerges that the
distribution of neutral gas is not a good tracer, and thus leads
to an incorrect LMC model. Instead, using the carbon star data,
they have provided an accurate measurement of the dynamical center
of the stars, which turns out to be consistent with the center of
the bar. Moreover, they have pointed out that the shape of the LMC
disk is not circular, but elliptical. In this case, the position
angle of the line of nodes is different from that of the major
axis of the LMC disk. These facts change completely the previous
geometry and a noticeable asymmetry between the two sides of LMC
divided by the line of the nodes arises when studying the
positions of the MACHO microlensing event towards LMC.
%
\section{Optical Depth and Asymmetry}
By using the van der Marel et al. description of LMC, we have
reported new accurate estimates of the optical depth for any
source/lens configuration [11]. Our results show an evident
near--far asymmetry of the optical depth values for lenses located
in the LMC halo. This asymmetry is due to the fact that the LMC
disk is inclined so that line of sights towards the far side of
LMC go through a larger portion of the LMC halo, according to the
prediction of Gould [12]. The flaring of the LMC disk is a further
source of the asymmetry of the optical depth between the far and
the near side of LMC. The asymmetry is completely lost, as
expected, in the self--lensing case.
%

We have also provided a set of statical calculations to study the
asymmetry among the locations of the microlensing events found.
Our results show that the existence of the near--far asymmetry
supports the common idea that the LMC is surrounded by a halo of
dark material [11].

%
\section{Modal mass estimates for lenses in LMC}
Evans and Kerins [7] have noticed that the distribution of
timescales of the events and their spatial variation across the
face of the LMC disk offers possibilities of identifying the
dominant lens population. We find that the observed timescales
distribution of the events found by the MACHO collaboration as a
function of the distance from the center of LMC is not in
agreement with that theoretically expected for lenses located all
in the inner galactic components of LMC. Our statistical analysis
allows to distinguish in principle at least two different lens
populations. In particular, only the events concentrated near the
center of LMC with self--lensing optical depth
$\tau_{\mathrm{sl}}\, >\, 2\times 10^{-8}$, have a good chance to
be true self--lensing events [11]. We have estimated the modal
values of the masses of these lenses. The results suggest that
they could belong to a low mass star population in the LMC
disk/bar components. In Fig. 1 we show the scatter plot of the
modal value of the lens mass $\mu$ with respect to the measured
Einstein time for each MACHO event. Filled (empty) triangles
represent points with $\tau_{\mathrm{sl}}\, >\, 2\times
10^{-8}\,(<\,2\times 10^{-8})$. The horizontal line indicates the
lower limit of the lens mass for self lensing. Four MACHO events
have a modal value of the lens mass smaller than the lower limit
and therefore are not represented in the figure. We also excluded
from our analysis the binary event [13] and the Galactic disk
event [14]. The dashed line represents the correlation line for
the $6$ events with $\tau_{\mathrm{sl}}\, <\, 2\times 10^{-8}$.
These events are located far from the LMC center and can hardly be
identified as belonging to a LMC population.
\begin{figure}[ht]
\begin{center}
\includegraphics[height=5cm,width=8cm]{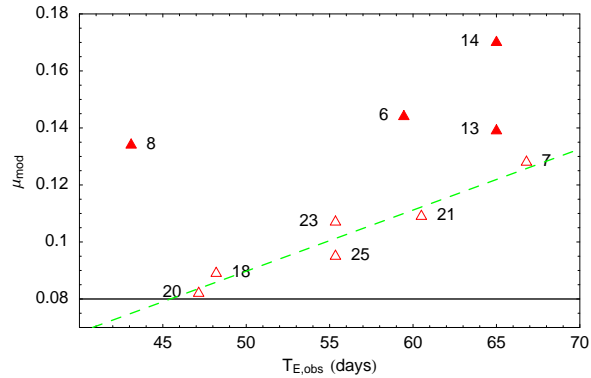}
\caption{\small Scatter plot of the modal value of $\mu$
($y$--axis) with respect to the measured Einstein time ($x$--axis)
of the event, where $\mu$ is the lens mass in solar mass units.
The label identifies the MACHO event as reported in [2].}
\label{tEVersusMu}
\end{center}
\end{figure}
\noindent
\vspace{-0.4cm}

\end{document}